# Interplay of spin and orbital ordering in the layered colossal magnetoresistance manganite La$_{2-2x}$Sr$_{1+2x}$Mn$_2$O$_7$ ($0.5 \leq x \leq 1.0$)

C. D. Ling*, J. E. Millburn, J. F. Mitchell, D. N. Argyriou and J. Linton

*Materials Science Division, Argonne National Laboratory, Argonne, Illinois 60439*

H. N. Bordallo

*Intense Pulsed Neutron Source, Argonne National Laboratory, Argonne, Illinois 60439*

# ABSTRACT

The crystallographic and magnetic phase diagram of the $n = 2$ layered manganite $La_{2-2x}Sr_{1+2x}Mn_2O_7$ in the region $x \geq 0.5$ has been studied using temperature dependent neutron powder diffraction. The magnetic phase diagram reveals a progression of ordered magnetic structures generally paralleling that of 3-D perovskites with similar electronic doping: A $(0.5 \leq x \leq 0.66) \rightarrow$ C $(0.75 \leq x \leq 0.90) \rightarrow$ G $(0.90 \leq x \leq 1.0)$. However, the quasi-2-D structure amplifies this progression to expose features of manganite physics uniquely accessible in the layered systems: (a) a "frustrated" region between the A and C regimes where no long-range magnetic order is observed; (b) magnetic polytypism arising from weak inter-bilayer magnetic exchange in the Type-C regime; and (c) a tetragonal to orthorhombic phase transition whose temperature evolution directly measures ordering of $d_{3y^2-r^2}$ orbitals in the $a$-$b$ plane. This orbital-ordering transition is precursory to Type-C magnetic ordering, where ferromagnetic rods lie parallel to the $b$-axis. These observations support the notion that $e_g$ orbital polarisation is the driving force behind magnetic spin ordering. Finally, in the crossover region between Type-C and Type-G states, we see some evidence for the development of local Type-C clusters embedded in a Type-G framework, directly addressing proposals of similar short-range magnetic ordering in highly-doped $La_{1-x}Ca_xMnO_3$ perovskites.

# I. INTRODUCTION

Colossal magnetoresistive (CMR) manganite perovskites have been intensively studied in recent years. The transport properties of these itinerant ferromagnets appear to be related to strong interactions among charge, spin and lattice degrees of freedom. Reducing the dimensionality of these perovskites by studying their behaviour in layered phases can help elucidate these interactions by constraining the lattice degrees of freedom and by enhancing the amplitude of charge magnetic fluctuations in the critical region above the insulator-metal transition. The $n = 2$ Ruddlesden-Popper (R-P) phase $La_{2-2x}Sr_{1+2x}Mn_2O_7$ (which can be written $SrO\bullet(La_{1-x}Sr_xMnO_3)_2$ to highlight the analogy to perovskites) has proved to be one of the most interesting such layered CMR manganites, exhibiting a remarkable range of magnetic behaviour.

Experimentally, the system $La_{2-2x}Sr_{1+2x}Mn_2O_7$ has been thoroughly described in the CMR region $0.3 \leq x \leq 0.5$, yielding an extraordinarily rich magnetic phase diagram. At low temperatures, a paramagnetic insulating (PI) state gives way to antiferromagnetic metallic (AFM)[1-3], ferromagnetic metallic (FM)[4-8], canted antiferromagnetic (CAF)[9] antiferromagnetic insulating (AFI)[4] and charge-ordered (CO)[5-8, 10] states. An AFI state was also found at $x = 1.0$[11, 12]. The intermediate region $0.5 < x < 1.0$, however, remained unexplored until very recently, when we were able to overcome the materials problems associated with this part of the phase diagram and successfully synthesise samples across its entire range[13].

In this paper, we present detailed neutron powder diffraction measurements on samples in the $Mn^{4+}$-rich half of the phase diagram. We find a progression of AFI phases [A $(0.5 \leq x \leq 0.66) \rightarrow$ C $(0.75 \leq x \leq 0.90) \rightarrow$ G $(0.90 \leq x \leq 1.0)$] generally following that of the 3-D perovskites, with notable differences resulting from the constraints of the quasi-2-D layered structure. In particular, we find: a region between the Type-A and Type-C stability fields where no long-range magnetic structure is stable; and a tetragonal-orthorhombic orbital ordering transition as well as magnetic polytypism in the Type-C regime. Furthermore, we present evidence for short-range Type-C magnetic structure motifs (*i.e.* ferromagnetic rods) embedded in a Type-G matrix in the crossover region between the stability fields of Type-G and Type-C. These observations directly address the hypothesis of similar phenomena in 3-D perovskites[14]. $MnO_6$ octahedral distortions are interpreted in terms of the occupancy, polarisation and ordering of $e_g$ orbitals and their relationship to magnetic ground state, suggesting that these are the driving forces behind spin ordering. The layered manganite structure thereby facilitates the observation of magnetic phenomena critical to understanding the relationship between spin and orbital degrees of freedom in manganites.

# II. EXPERIMENTAL

Synthesis of samples in the range $0.5 \leq x \leq 1.0$ followed the method described previously[13]. All samples were annealed to completely fill oxygen vacancies present in the quenched samples. Oxygen contents were verified by thermogravimetric analysis.

Preliminary synchrotron x-ray powder diffraction (XRD) data were collected on beamline X7A at Brookhaven National Laboratory's National Synchrotron Light Source. Subsequent temperature-dependant synchrotron XRD data were collected for certain samples ($x = 1.00, 0.95, 0.90, 0.80, 0.75, 0.55, 0.50$) on beamline 12-BM-B (BESSRC-CAT) at Argonne National Laboratory's Advanced Photon Source (APS). Full-width-half-maxima (FWHM) for the (2 0 0) and (0, 0, 10) reflections at 300 K ($\lambda = 0.688$ Å, step size 0.005 °2$\theta$) are listed in Table 1, and compare favourably with the $Al_2O_3$ calibration standard used (FWMH = 0.035(5) °2$\theta$). These results give us additional confidence in the crystallinity and chemical homogeneity of our samples, bearing in mind the potential for misleading results obtained with inhomogeneous samples, as discussed by Battle *et al.*[15]. A slight broadening of (2 0 0), but not (0, 0, 10), for $0.75 \leq x < 0.95$ will be discussed in Section III.C.

Temperature-dependant time-of-flight (TOF) neutron powder diffraction data were collected on the Special Environment Powder Diffractometer (SEPD), the High Intensity Powder Diffractometer (HIPD) and the General Purpose Powder Diffractometer (GPPD) at Argonne National Laboratory's Intense Pulsed Neutron Source (IPNS). Data were analysed using the program suite GSAS[16]. The perovskite $La_{1-x}Sr_xMnO_3$ was included in a number of refinements as a minor impurity, between $0 \sim 3$ wt%. No evidence was found for symmetry lowering below *I4/mmm* with the exception of the orthorhombic regime discussed in Section III.C. The difference between the neutron scattering lengths of La and Sr was insufficient to adequately assess possible ordering on mixed La/Sr sites, therefore these sites were treated as disordered mixtures of La and Sr in ratios appropriate to the value of $x$. Oxygen sites were fixed at 100 % occupancy in accordance with thermogravimetric analyses. Unit cells, atomic positions, isotropic thermal displacement parameters, absorption and extinction of the samples were refined in addition to the background and peak profiles of the diffraction patterns.

# III. RESULTS

The magnetic phase diagram of $La_{2-2x}Sr_{1+2x}Mn_2O_7$ across the currently accessible range of the solid solution, $0.3 \leq x \leq 1.0$, is presented in Fig. 1. A number of magnetic phases are identified: AFM, FM, A, C, C*, G. Diagrams of the spin arrangements for each of these phases are presented in Fig. 2; detailed descriptions of these structures can be found in the text. In this section, we will describe the various stability fields in the phase diagram and comment on the influence of crystallographic structure on the magnetic ground state.

The region $0.3 \leq x \leq 0.5$ has been studied in some detail by several groups. At room temperature, the entire range is PI. At $x = 0.3$ the material is AFM below the insulator-metal transition temperature[3] $T_{IM} \approx 100$ K [Fig. 2(a)]. A tilted FM state is found below the Curie temperature $T_C$ in the region $0.32 \leq x \leq 0.36$, while in-plane ferromagnetism is found between $0.36 \leq x \leq 0.40$[4-8] [Figs. 2(b-c)]. The region $0.42 \leq x \leq 0.50$ is Type-A AFI below the Néel temperature $T_N$[4] [Fig. 2(d)]. A lower temperature transition in the range $0.42 < x < 0.48$ gives rise to a CAF ground state [Figs. 2(c-d)]. At $x = 0.50$ a CO state appears below $T_{CO}$[5-8, 10], which is slightly above $T_N$, below which the Type-A AFI state replaces it[17]. At the far end of the phase diagram, $x = 1.00$ is Type-G AFI below $T_N$ [Fig. 2(h)]. Magnetic states in the newly elucidated region $0.50 < x < 1.00$ are described in the following subsections.

## A. Type-A AFI ($0.42 < x < 0.66$)

As $x$ is increased beyond the CO composition $x = 0.50$, we do not observe Bragg peaks due to long-range CO in synchrotron XRD data. Reflections due to the Type-A AFI state in neutron powder diffraction persist to $x = 0.65$. For $x$ close to 0.50, the refined Type-A magnetic moment $\mu_A$ at 20 K is close to the expected spin-only moment of 3.5 $\mu_B$/Mn, indicating a highly ordered spin state (Fig. 3). As $x$ increases, $\mu_A$ decreases slowly up to $x = 0.60$ before dropping rapidly to zero when $x = 0.66$. A typical low-temperature magnetic refinement ($x = 0.58$ at 20 K, $\mu = 2.32(5)$ $\mu_B$/Mn) is shown in Fig. 4, with corresponding data in Table 2. Perovskite-type bilayers in the Type-A AFI state [Fig. 2(d)] consist of ferromagnetic sheets perpendicular to $c$, with antiferromagnetic intra-bilayer coupling and ferromagnetic inter-bilayer coupling. No diffraction evidence was found for canting of the Type-A spins when $x > 0.50$, either within or between ferromagnetic sheets.

## B. Frustrated spin state ($0.66 \leq x \leq 0.74$)

In samples of composition $x = 0.66$, 0.68 and 0.70, no evidence for any kind of long-range magnetic order was observed in neutron powder diffraction data down to 20 K, despite an available spin-only moment in the order of 3.3 $\mu_B$/Mn. Fig. 3 highlights the sharp decrease in the proportion of long-range ordered spins on either side of this region, indicating the presence of a frustrated spin state. The possible existence of a spin-glass in this region will be investigated in the near future using frequency-dependent magnetic susceptibility and/or specific heat measurements.

## C. Type-C/C* AFI ($0.74 < x < 0.92$)

A crystallographic phase transition was observed on cooling samples in the range $0.75 < x \leq 0.94$ (Fig. 1). Single diffraction peaks such as (2 0 0) split cleanly into triplets, indicating a tetragonal $\rightarrow$ orthorhombic phase transition, *i.e.* the triplet is $(2\ 0\ 0)_T + (2\ 0\ 0)_O + (0\ 2\ 0)_O$ (Fig. 5). No reflections violating body-centering extinction conditions in the orthorhombic phase were observed in neutron powder diffraction data or synchrotron XRD data, leaving *Immm* as the only appropriate (orthorhombic) maximal non-isomorphic subgroup of *I4/mmm*. Rietveld-refinement of thermal displacement parameters provided no justification for further symmetry lowering (*i.e.* breaking of mirror symmetries), therefore *Immm* was retained as the orthorhombic space group. The magnitude of the orthorhombic splitting (at all temperatures) maximises at $x = 0.80$. For $x = 0.80$, the refined *I4/mmm* crystal structure at room temperature is compared to the *Immm* crystal structure at 10 K in Table 2. As discussed below, this crystallographic phase transition is associated with an orbital ordering transition that preferentially orients the occupied $Mn^{3+}$ $e_g$ orbital along the $y$-axis. Observation of all three $\langle 2\ 0\ 0 \rangle$ peaks indicates that tetragonal and orthorhombic phases coexist to T = 10 K, reflecting the first order nature of this orbital ordering transition.

In approximately the same composition range ($0.75 \leq x \leq 0.90$) (Fig. 1), a new set of AF diffraction peaks appears below $T_N$. These peaks can be divided into two subsets (Fig. 5), the relative intensities of which vary (apparently unsystematically) with $x$. The more intense subset can be indexed in terms of the *Immm* nuclear subcell to ($h+1/2, k, l+1/2$) and the weaker subset to ($h+1/2, k, l$). This indicates the presence of two distinct but related AF superstructures. The doubling of the short $a$-axis, but not the long $b$-axis, in both cases indicates the presence of ferromagnetic columns along $b$, antiferromagnetically coupled along $a$. Comparing calculated to observed magnetic intensities clearly indicates that these columns are also antiferromagnetically coupled along $c$ within the perovskite-type bilayers. This is analogous to the Type-C AFI state first seen in $La_{1-x}Ca_xMnO_3$, $x \approx 0.8$[18] and more recently in $Sm_{1-x}Ca_xMnO_3$ and $Pr_{1-x}Sr_xMnO_3$, $x \approx 0.85$[19]. Magnetic transitions in the latter compounds were also accompanied by structural transitions. Inter-bilayer coupling is neither ferromagnetic nor antiferromagnetic, consistent with the degeneracy in the space group symmetry *Immm*; however, coupling between one bilayer and the next-nearest bilayer is not degenerate, and the two possibilities available are the source of the two subsets of AF reflections, *i.e.* there are two magnetic polytypes. The minority phase is hereafter referred to as Type-C [Fig. 2(e)], and the majority phase, in which *c* is doubled, as Type-C* [Fig. 2(f)]. Rietveld-refinement indicates that the spins are aligned parallel to the long basal plane axis, *b*. Two-phase magnetic refinements were carried out on the assumption that $\mu_C = \mu_{C*}$. Final Rietveld-refined neutron powder diffraction data ($x = 0.84$) are presented in Fig. 5, and key results ($x = 0.80$) in Table 2.

The presence of well-defined magnetic reflections for both Type-C and Type-C* in neutron powder diffraction data rules out the possibility that the minority Type-C phase represents stacking faults in Type-C* rather than a separate phase. Although stacking faults could be expected when the interaction differentiating the two phases takes place across approximately 16 Å, only long-range-ordered Type-C regions can account for the observed data. Stacking fault intergrowths, such as those seen at $x \approx 0.4$ by Osborn *et al.*[20], only lead to a diffuse scattering streak parallel to the *c*-axis. Gurewitz *et al.*[21] reported an analogous form of magnetic polytypism in the $n = 2$ R-P phase Rb$_3$Mn$_2$Cl$_7$, where individual bilayers have a Type-G AFI arrangement. They found that different methods of single-crystal growth led to the formation of different magnetic polytypes. We have as yet been unable to grow a single crystal of La$_{2-2x}$Sr$_{1+2x}$Mn$_2$O$_7$ within the Type-C/C* AFI regime with which to further investigate this magnetic polytypism.

The tetragonal $\rightarrow$ orthorhombic phase transition was observed to be incomplete, even at 10 K. Consequently, in order to obtain meaningful values for the Type-C/C* magnetic moment, it was necessary to determine whether the Type-C/C* state resides in one or both crystallographic phases. Neutron powder diffraction data at $x = 0.80$ were collected on both cooling and warming over the range $10 \leq T \leq 300$ K. The refined tetragonal crystallographic phase fraction is shown on the left-hand axis of Fig. 6(a). At 300 K, 14.0(9) % is already orthorhombic, indicating that the transition temperature $T_O > 300$ K, although the bulk of the sample remains tetragonal. An approximate value of $T_O \approx 340$ K can be determined by extrapolating lattice parameters [Fig. 11(c)] to the mutual intersection of $a_O$, $b_O$ and $a_T$. The orthorhombic phase fraction remains temperature independent down to $T \approx 200$ K, below which the transformation rapidly accelerates. At $T \approx 90$ K, only 6.9(2) % remains tetragonal, and this phase fraction persists down to 10 K. The transformation was reversible on warming, with a hysteresis of approximately 30 K. The experiment was conducted twice, using two different cooling/warming rates (5 and 0.1 Kmin$^{-1}$); both the hysteresis and the final low-T tetragonal phase fraction were found to be rate-independent in this range. Only Type-C* (not Type-C) magnetic reflections were seen in low-T data for the $x = 0.80$ sample. The right-hand axis of Fig. 6(a) shows the refined Type-C* magnetic moment, calculated on the assumption that the tetragonal and orthorhombic phase fractions are equally responsible for these reflections. The hysteresis in the magnetisation has the form of a magnetisation curve convoluted with the crystallographic hysteresis, suggesting that Type-C* exists in the orthorhombic phase fraction only. Recalculating $\mu_{C*}$ on this basis gives a magnetisation curve with no hysteresis [Fig. 6(b)], confirming that the Type-C* magnetic superstructure only exists in the presence of an orthorhombic subcell. Note that for $x = 0.80$ and $x = 0.82$ (Table 3), the orthorhombic:tetragonal crystallographic ratios are similar but the C:C* magnetic phase ratios are very different; this rules out the possibility that Type-C resides in the tetragonal phase fraction while Type-C* resides in the orthorhombic phase fraction. Note also that for $x = 0.80$, the observed moment is only 60 % of the expected spin-only moment (Fig. 3), indicating that a considerable fraction of the Mn spins do not take part in long-range Type-C* ordering.

This approach to the refinement of SEPD neutron powder diffraction data was used to determine crystallographic and magnetic phase fractions and magnetic moments for other samples in the Type-C/C* regime. Final results are presented in Table 3 and Fig. 3. Note in Table 3 that for all samples where the crystallographic phase fraction could be refined, a small proportion of the sample remains tetragonal at 20 K. Similarly, a small proportion of these samples is found to be orthorhombic at room temperature. The behaviour shown for $x = 0.80$ in Fig. 6(a) cannot, therefore, be attributed to compositional inhomogeneity, although it may arise from other small inhomogeneities within the samples such as in crystallinity or oxygen stoichiometry. This behaviour cannot be attributed to compositional inhomogeneity, as evidenced by the FWHM of synchrotron XRD reflections (Table 1); there is a slight broadening of (2 0 0) in the orthorhombic regime but no broadening of (0, 0, 10), indicating that a small orthorhombic component rather than compositional inhomogeneity [as in Fig. 7] causes the broadening.

For samples at the lower and upper limits of the Type-C/C* regime, certain reasonable assumptions were necessary to obtain the values reported in Table 3. The result of these assumptions is a consistent picture of the composition dependence of the crystallographic and magnetic phase diagram. At $x = 0.75$, a Type-C/C* magnetic moment (Fig. 3) could be refined, but no orthorhombic splitting [Fig. 7] was observed (even with synchrotron XRD data) *i.e.* the subcell appears to remain metrically tetragonal (Fig. 1). Future electron microscopy studies might reveal some twinning phenomena which help to elucidate this point; for the purposes of the present study, a single phase tetragonal model was assumed. The refined magnetic moment for $x = 0.75$ given in Table 3 has been calculated on the assumption that if Type-C/C* only requires an infinitesimal orthorhombicity at this composition, then the entire sample will be in the Type-C/C* state below $T_N$.

For $x = 0.90$ at low temperatures, crystallographic phase fractions for the $I4/mmm$ and $Immm$ components were refined to give 94 wt% orthorhombic and 6 wt% tetragonal. However, determination of $\mu_{C/C^*}$ was complicated by the simultaneous presence of Type-G and Type-C* magnetic reflections at low temperatures. Type-C reflections were not observed; however, considering the typically small ratio of Type-C:Type-C* reflections throughout the Type-C/C* regime (*e.g.* at $x = 0.84$, Fig. 5), and the weakness of Type-C* reflections for $x = 0.90$, Type-C reflections may be simply too weak to observe at this composition. The possibility that the simultaneous presence of Type-G and Type-C* reflections is due to compositional inhomogeneity is remote, based on analysis of the FWHM of synchrotron XRD peaks (Table 1) in Section II. Initial refinements of the magnetic reflections assumed that Type-G ordering arose only in the tetragonal and Type-C/C* ordering only in the orthorhombic phase fractions. These assumptions were based on our observations of neighbouring compositions. This model produced an unphysically high value for $\mu_G = 28 \mu_B$/Mn, indicating that the observed Type-G intensities could not arise from a mere 6 wt% tetragonal phase fraction. Some of the orthorhombic phase therefore also exhibits Type-G magnetic ordering.

Since the Type-G and Type-C/C* magnetic states both exist in the orthorhombic crystallographic phase, it is impossible to deconvolute the magnetic from the crystallographic phase fractions *i.e.* the refinement of $\mu_{C/C^*}$ and $\mu_G$ is underdetermined. Stable refinements can only be carried out when certain assumptions are made about the relationship between these magnetic phases. There are two end-case models, which are assessed below against the plausible criterion that the moments within the Type-C/C* and Type-G domains of the $x = 0.90$ sample are smoothly continuous with the moments across the whole of the Type-C/C* and Type-G regimes (as a function of $x$) (Fig. 3). The relationship between the refined magnetic moments and the intensities of observed magnetic reflections in these models is fully explained in the Appendix. Analysis of these two models leads to important conclusions regarding the evolution of magnetic ordering as a function of $x$.

In the first model, long-range-ordered Type-C/C* regions exist in different crystallites (of a polycrystalline sample) to those which contain long-range-ordered Type-G regions. This is a *chemical phase segregation* model, consistent with phase segregation due to compositional inhomogeneity. This model does not remove the additional degree of freedom from the refinement (Eqn. 9 in the Appendix), therefore it is necessary to fix either the crystallographic or the magnetic phase ratio. The most obvious constraint is $\mu_{C/C^*} = \mu_G$ (Eqn. 10). The refined magnetic moment obtained using this constraint is shown as an open marker in Fig. 3, where it can be seen that this model results in an unexplainable upturn of the $\mu_{C/C^*}$ *vs.* $x$ curve. From Eqn. 9, it can be seen that increasing the C/C* orthorhombic phase fraction $^C F_o$ causes $\mu_{C/C^*}$ to decrease towards its value at $x = 0.88$, but $\mu_G$ to increase towards its value at $x = 0.92$. There is in fact no value of $^C F_o$ for which neither the $\mu_{C/C^*}$ nor $\mu_G$ *vs.* $x$ curve displays an anomalous upturn at $x = 0.90$. This model is therefore rejected. Note that the rejection of this model is further evidence that compositional inhomogeneity is not a significant factor in our samples, and hence is not impacting our phase diagram.

In the second model, long-range-ordered Type-C/C* regions and long-range-ordered Type-G regions exist in the same crystallites (of a polycrystalline sample). This is a *magnetic phase segregation* model that does not involve chemical phase segregation. In this case, the refinement is no longer underdetermined (Eqns. 12a and 12b in the Appendix). The results are shown as grey markers in Fig. 3, which fall on the $\mu_{C/C^*}$ and $\mu_G$ *vs.* $x$ curves smoothly extrapolated from the rest of the Type-C/C* and Type-G regimes. Clearly, while the true nature of the phase segregation at $x = 0.90$ may fall between the two end-case models considered, the second (magnetic phase segregation) is a far better approximation based on available data. It should still be remembered, however, that the refined moments for $x = 0.90$ (Fig. 3, Tables 2 and 3) are necessarily less certain than those for the rest of the Type-C/C* and Type-G regimes.

In contrasting the two models for $x = 0.90$, it is important to note the difference between the sets of magnetic moment terms which *do not* contribute to observed neutron diffraction intensity, *i.e.* the terms subtracted from $\mu_{TOT}$ on the RHS of Eqn. 8 (chemical phase segregation model) and Eqns. 11a and 11b (magnetic phase segregation model). The RHS of Eqn. 8 has the same form as the RHS of Eqns. 4 and 5, which refer to the Type-G and Type-C/C* regimes, where the only terms not contributing to observed intensity are non-long-range-ordered (subscript D) terms. On the RHS of Eqn. 11a, however, a long-range-ordered term ($\mu_{oC}$) is also subtracted from $\mu_{TOT}$. If the composition ($x$) at the boundary between Type-G and Type-G+C/C* regimes is X, then at a slightly higher composition X + $\delta$, $\mu_{oC} = 0$ and Eqn. 11a $\equiv$ Eqn. 4 (Type-G). As $\delta \rightarrow 0$ and then becomes negative, $\mu_G$ must decreases smoothly through the phase boundary in this model (Fig. 3) (a discontinuity in $\mu_G$ would imply the alternative, chemical phase segregation, model). Therefore, since $\mu_{C/C^*}$ becomes non-zero discontinuously at the first-order phase transition when $\delta = 0$, it must do so at the expense of $\mu_{oD}$; *i.e.*, regions of Type-C/C* grow out of non-long-range-ordered orthorhombic regions, rather than out of long-range-ordered Type-G regions, and the total number of long-range-ordered spins increases. This can be seen in the result that ($\mu_G + \mu_{C/C^*}$) at $x = 0.90$ is greater than $\mu_G$ at $x = 0.92$ (Fig. 3, Tables 3 and 4). Analogous arguments hold when the Type-G+C/C* regime is entered from the Type-C/C* regime. The magnetic phase segregation model therefore has important implications for the nature of non-long-range-ordered spins in the Type-C/C* and Type-G regimes, to be discussed further in Section IV.

The temperature dependence of $\mu_{C/C*}$ and of $\mu_G$ at $x = 0.90$, normalised to their 10 K values *i.e.* independent of a model for phase separation, are shown in Fig. 8. Because it is determined by a single point, at this stage we can attach no significance to the local minimum in T = 60 K in the $\mu_{C/C*}$ curve. However, that $\mu_{C/C*}$ decreases below 35 K while $\mu_G$ is increasing demonstrates that the magnetic phases are in competition within the same crystallites. This behaviour is consistent with the second model for the phase separation, and further demonstrates that the observation of both magnetic phases is not a result of compositional inhomogeneity.

### D. Type-G AFI (0.89 < $x$ ≤ 1.00)

Low temperature neutron powder diffraction data at $x = 0.90$ display a number of magnetic peaks in addition to those of the Type-C/C* AFI. These peaks can be indexed to the $I4/mmm$ and/or $Immm$ subcells as ($h$+1/2 $k$+1/2 $l$), indicative of the Type-G AFI state found at $x = 1.00$[11, 13]. At $x = 0.92$, only Type-G reflections remain. Type-G reflections persist to $x = 1.00$. The relative intensities of Type-G reflections vary as a function of $x$ over the whole range $0.90 \leq x \leq 1.00$, however, Type-C/C* reflections are not observed for $x > 0.90$. The low-temperature orthorhombic transition seen in the Type-C/C* regime persists as far as $x = 0.95$ (Fig. 1). For $0.90 \leq x \leq 0.94$, the magnitude of the splitting was sufficient to refine orthorhombic lattice parameters but insufficient to refine a tetragonal/orthorhombic phase fraction from SEPD data. At $x = 0.96$, no splitting of the (2 0 0) reflection could be observed, even in synchrotron XRD data.

The ideal Type-G state at $x = 1.00$ involves fully AF coupling between nearest neighbour Mn sites, with all spins parallel to $c$. Since no additional magnetic reflections appear, the only available degree of freedom in the Type-G model with which to fit SEPD data across the entire regime was to apply a uniform, parallel, tilting of the spins away from $c$ and into the $a$-$b$ plane. This tilting is defined by $\theta$, the polar angle of $\mu$ to $c$. The direction of the $a$-$b$ component of the spin vector (defined by $\phi$, the azimuthal angle of $\mu_{xy}$ to $a$) cannot be determined in the tetragonal space group $I4/mmm$ ($x > 0.95$), therefore the diagonal between $a$ and $b$ ($\phi = 45$ °) is chosen arbitrarily. As the orthorhombic splitting increases below $x = 0.95$, differentiating $a$ from $b$, refinement of $\phi$ is in principle possible. At $x = 0.92$, $\phi$ refines to a stable value of 65(7) °, *i.e.* it tends towards the longer $b$ axis as observed in ordered Type-C/C*. Representative 20 K refinements ($x = 0.98$ and 0.92) are shown in Fig. 9, with corresponding data in Table 2. Table 4 lists the final refined spin magnitudes and orientations as they evolve across the Type-G AFI regime, the magnitudes also being plotted in Fig. 3. The treatment of $x = 0.90$ data was described in Section III.C. Note that at $x = 0.90$, Type-G reflections were too weak to refine the angle azimuthal to $a$; $\phi$ was fixed at 45 °, which was found to give a better refinement than the $x = 0.92$ value of 65 °. Trends in spin orientation ($\theta$ and $\phi$ in Table 4) therefore suggest that the Type-G magnetic state when $x = 0.90$ is closer to ideal Type-G ($x = 1.00$) than when $x = 0.92$. The interpretation of trends in $\theta$ and $\phi$, including their apparent reversal when $x < 0.95$, will be discussed further below in terms of the mixed Type-G+C/C* state described above for $x = 0.90$.

## IV. DISCUSSION

The composition- and temperature-dependence of lattice constants and Mn-O bond lengths can illuminate details about electronic states of the layered manganites. $MnO_6$ octahedra do not rotate in these layered phases as they do in the perovskite analogues (hence the _mmm_ symmetry), therefore $a$ and $c$ are directly related to the equatorial and axial Mn-O bond lengths respectively. Changes in the relationship between equatorial and axial bond lengths can be interpreted in terms of the distribution of $e_g$ electron density associated with $Mn^{3+}$ ions $(1 - x)$ into either planar $d_{x^2-y^2}$–type or linear $d_{3z^2-r^2}$–type orbitals. The relative stability of these orbitals is thought to play an important role in determining the ground state magnetic structure of both 3-D[22] and layered[23] manganites. Maezono and Nagaosa[24] have produced a theoretical phase diagram for $n = 2$ manganites based on orbital polarisation, Coulombic repulsion and coupling with lattice deformation. Extended into the high-$x$ region, this model predicts a smooth transition from Type-A to Type-G over the range $0.5 \leq x \leq 1.0$. Having accessed this region experimentally and found the magnetic phase diagram to be somewhat different, we are now in a position to comment further on the relationship between orbital polarisation and magnetic structure. In considering this relationship on the basis of a neutron powder diffraction study, it should be remembered that while this technique is far more sensitive to spin ordering than to orbital ordering, it is not completely insensitive to the latter. Consequently, the absence of direct evidence for orbital ordering is not a conclusive result in itself, and orbital ordering patterns discussed below are inferred from (directly observed) spin ordering patterns. We will also comment on (limited) evidence for an extension of the CO regime as far as $x = 0.6$.

Trends in room temperature lattice parameters for $x > 0.5$ are a reversal of those for $x < 0.5$[3, 25] [Fig. 7]. The $a$-axis reaches a maximum at $x \approx 0.5$, and the minimum in $c/a$ is only slightly offset from this point by a minimum in $c$ at $x \approx 0.6$. Lattice parameters at low temperatures (10 - 20 K) track those at room temperature [Fig. 7], with the obvious exception of the region $0.75 < x < 0.95$ where the bulk of the sample is orthorhombic at low temperatures. $MnO_6$ octahedral bond lengths also vary smoothly with composition (Fig. 10). Close to the magnetic transition temperatures, however, the behaviour of lattice parameters varies significantly between compositions (Fig. 11). Coherent lattice effects such as these have been considered in detail elsewhere for $x < 0.5$[3, 25-29]; for $x \geq 0.5$, they are considered below in terms of qualitative rules of orbital exchange interactions and Jahn-Teller (J-T) distortions of $Mn^{3+}O_6$ octahedra.

The CO state seen at $x = 0.50$ has analogues in many perovskite phases $Ln_{1-x}Ca_xMnO_3$ when $x = 0.50$[30-36]. These phases adopt a Type-CE AFI structure below $T_N < T_{CO}$. The gap between $T_{CO}$ and $T_N$, _i.e._ the relative thermodynamic stability of AFI _vs._ CO states, is very sensitive to both $x$ (nominal hole concentration) and average cation radius on the perovskite $A$-site[30]. In perovskites, Akimoto _et al._[22] related these sensitivities to the one-electron bandwidth $W$: larger $A$ cations lead to straighter Mn–O–Mn angles and thus larger $W$, delocalising $d_{x^2-y^2}$ orbitals and facilitating ferromagnetic double-exchange in the $a$-$b$ plane. Smaller $A$ cations lead to smaller $W$, localising $e_g$ electrons and favouring a CO (Type-CE AFI) state. Increased hole doping ($x$) further facilitates double-exchange.

In the layered manganites, however, $MnO_6$ octahedra do not tilt as in perovskites, so the effect of cation size on $W$ is less obvious or pronounced. Nonetheless, variations in lattice parameters (Fig. 7) and Mn-O bond lengths (Fig. 10) with $x$ in the Type-A AFI regime are consistent with Akimoto _et al._'s model[22] and with ideas from Maezono and Nagaosa[24] about planar orbital stabilisation. Note that since $x = 1.00$ implies that all Mn ions are in the non-J-T active 4+ oxidation state, the distortion of $MnO_6$ octahedra at $x = 1.00$ can be taken as the baseline distortion due solely to the steric influence of surrounding cations. Octahedra in the Type-A regime are distorted from this baseline by a contraction of the axial bond Mn-O1 (within the perovskite-type layer), shortening Mn-Mn distances along $c$. This short Mn-Mn distance is the one over which an antiferromagnetic superexchange interaction takes place in Akimoto *et al.*'s model for Type-A. The ferromagnetic double-exchange interaction within the *a-b* plane is less distance dependant, hence the Type-A regime is centred around the minimum in $c/a$ [Fig. 7]. As $x$ increases, $e_g$ electron density is lost, weakening the double-exchange interaction and causing $c/a$ to increase again (back towards the non-J-T distorted value at $x = 1.00$), both effects contributing to the disappearance of the Type-A state at $x = 0.66$.

The presence of antiferromagnetically coupled ferromagnetic planes in Type-A suggests anisotropic transport properties, and perhaps foreshadows metallic conductivity in the *a-b* plane given the picture of Akimoto *et al.*[22] and Maezono and Nagaosa[24] of mobile $e_g$ electrons in a $d_{x^2-y^2}$ band mediating double-exchange. In the system $(La_{1-z}Nd_z)_{1-x}Sr_xMnO_3$, where the additional variable $z$ controls average A-site radius, Akimoto *et al.*[22] did in fact find (weak) metallic conductivity in the Type-A regime ($x = 0.54$) when $z = 0.0$. As $z$ was increased, resistivity increased, but the general behaviour remained metallic up to $z = 0.8$ except at low temperatures ($\leq 60$ K) where a weak localisation effect was observed. In our ceramic samples, normalised resistance $R/R_{295K}$ *vs.* T plots for various values of $x$ within the Type-A regime are shown in Fig. 12. Although the resistivity is much too high to justifiably call these materials metallic (one must, of course, consider grain boundary effects not present in single crystal samples), for $x \leq 0.60$ there is a slight downturn in $R/R_{295K}$ below $T_N$. We note that for $0.5 < x < 0.6$, $R/R_{295K}$ shows a small anomaly around $T_N$. This anomaly is similar to that observed in in-plane resistivity measurements of single crystals at the CO composition $x = 0.50$[6, 10]. Unfortunately, neutron powder diffraction and synchrotron XRD have been unable to detect Bragg peaks associated with long-range CO in this region, or diffuse scattering in the case that CO is only short-range. Electron microscopy studies are underway on the polycrystalline samples, and a future single-crystal study searching for CO diffraction peaks may resolve the issue.

In the region $0.50 \leq x \leq 0.55$, significant coherent lattice effects are observed as a function of T. For example, upon cooling the $x = 0.55$ sample through $T_N$, there are anomalies in the temperature dependence of the both the $c$ and $a$ axes similar to those observed for $x = 0.50$[17] [Fig. 11(a)]. The change in slope in the temperature dependence of $c$ close to $T_N$ is consistent with transfer of charge into $d_{x^2-y^2}$ orbitals below $T_N$[17, 37]. A similar trend (much smaller in magnitude) is observed when $x = 0.58$ [Fig. 11(b)]. The small anomaly observed in $a$ for $x = 0.55$ (a small increase) [Fig. 11(a)] has also been reported for layered manganites with lower $x$ and may be ascribed to magnetostrictive effects[29].

The Type-C/C* AFI state is analogous to the Type-C state in $La_{1-x}Ca_xMnO_3$[18]. Goodenough[38] described this magnetic structure at $x = 0.75$ using qualitative rules of semi-covalent exchange, with "Case 1" (filled-filled *d*-orbital) antiferromagnetic interactions taking place in a plane and "Case 2" (filled-empty *d*-orbital) ferromagnetic interactions taking place perpendicular to that plane. Although one in four ferromagnetic interactions (one in twelve of all interactions) are unfavourable, this is the most cooperative possible arrangement of localised $d_{3y^2-r^2}$ orbitals at this composition. The proposed orbital ordering pattern and spin orientations in the *a-b* plane are shown in Fig. 13(d). There is no diffraction evidence for such orbital ordering, and the range of $Mn^{3+}$ doping across the Type-C/C* solid-solution (10-25 % of Mn sites) rules it out as an exact pattern. Nonetheless, lattice effects (*i.e.* orthorhombic splitting) do clearly indicate that $e_g$ electron density from $Mn^{3+}$ ions is concentrated along $b$. This suggests a model in which $d_{3y^2-r^2}$ orbitals are delocalised along $b$ to form bands [dashed lines in Fig. 13(d)]. These $d_{3y^2-r^2}$ bands mediate ferromagnetic double exchange while AF superexchange interactions dominate perpendicular to them. This is analogous to the model proposed by Martin *et al.*[19] for Type-C ordering in the perovskites $Sm_{0.15}Ca_{0.85}MnO_3$ and $Pr_{0.15}Sr_{0.85}MnO_3$. Note that in the present case, an orthorhombic lattice distortion could be avoided by placing the ferromagnetic rods along $c$ rather than $b$. That the distortion nonetheless takes place supports the notion that ferromagnetic coupling in Type-C/C* requires long-range delocalisation, which cannot occur along $c$ in the layered phase. Transport measurements on polycrystalline samples reveal that all compounds in this region of phase space are highly insulating. If suitable untwinned single crystals could be prepared, they should reveal anisotropic conductivity in the *a-b* plane if this picture is correct.

The present model for Type-C/C* implies that orbital ordering along $b$ (and hence symmetry lowering) is a precondition for spin ordering, *i.e.* $T_O > T_N$. Fig. 1 supports this hypothesis, $T_O$ tracking approximately 150 K above $T_N$ as $x$ varies. As discussed in Section III.C, however, between $T_O$ and $T_N$ the bulk of the sample remains tetragonal, *i.e.* the tetragonal phase exists in a metastable, supercooled form. This behaviour is observed across the whole of the Type-C/C* regime, and therefore cannot be explained by gross phase segregation into Type-C/C* and neighbouring regimes due to compositional inhomogeneity. The proportions of crystallites which transform immediately at $T_O$ or remain untransformed at 10 K vary unsystematically with $x$, and may reflect small inhomogeneities in factors such as crystallinity or oxygen stoichiometry (although small compositional inhomogeneities may play some role, they cannot account for this behaviour across the whole regime). The hysteresis was observed to be rate-independent. It is interesting to note the coincidence of the crystallographic hysteresis loop with $T_N$ when $x = 0.80$ [Fig. 6(a)], suggesting a possible relationship between them. Structural phase transitions were also noted to coincide with $T_N$ in the Type-C perovskites $Sm_{0.15}Ca_{0.85}MnO_3$ and $Pr_{0.15}Sr_{0.85}MnO_3$[19]. The simplest explanation is that for $T_N < T < T_O$, the energy difference between tetragonal and orthorhombic states is insufficient to drive the transition in the bulk of the crystallites; then for $T < T_N$, the potential for spin ordering in the orthorhombic state increases the energy difference and drives the transformation for the bulk of the crystallites. However, the temperature at which the bulk $T \rightarrow O$ transition occurs (*i.e.* the steepest part of the hysteresis curve in Fig. 6(a), as opposed to the actual $T_O$) does not track $T_N$ as $x$ increases from 0.80, but in fact remains almost $x$ invariant. Any relationship between $T_N$ and the bulk transition must therefore be more subtle, possibly involving short-range spin ordering, the onset of which does not necessarily track $T_N$ and is not visible to neutron powder diffraction.

The Type-G AFI regime is also analogous to the magnetic structure of $La_{1-x}Ca_xMnO_3$ over the same range of $Mn^{3+}$ doping $x$[18]. As explained by Goodenough[38], Type-G at $x = 1.00$ is a consequence of AF superexchange interactions between all nearest neighbour $Mn^{4+}$ sites. Isolated $Mn^{3+}$ sites, introduced into the Type-G lattice as $x$ decreases from 1.00, have ferromagnetic double-exchange interactions with their $Mn^{4+}$ neighbours. This reduces the total AF moment without creating a net ferromagnetic moment, in agreement with observations of a smoothly decreasing moment as $x$ decreases from 1.00 (Fig. 3).

There are, however, a number of indications that this simple model for Type-G as $x$ decreases from 1.00 is incomplete. Firstly, Type-G spins appear to 'tilt' into the *a-b* plane (Table 4). Secondly, with decreasing $x$ the equatorial Mn-O3 bonds lengthen while axial Mn-O1 and Mn-O2 bonds remain essentially constant (Fig. 10), indicating that $e_g$ electron density from doped $Mn^{3+}$ ions is concentrated in the *a-b* plane. Thirdly, the tetragonal-orthorhombic phase transition associated with Type-C/C* extends from $x \approx 0.75$ to $x \approx 0.95$ (Fig. 2), indicating that in this region, $e_g$ electron density is concentrated along *b*. All these features suggest a transition towards Type-C/C* as $x$ decreases from 1.00, however, the absence of Type-C/C* neutron diffraction reflections rules out a continuous long-range ordered canting of spins from Type-G to Type-C/C*.

A qualitative picture explaining these observations is illustrated in Fig. 13. Isolated $Mn^{3+}$ sites are introduced into the Type-G framework [Fig. 13(a)], but by virtue of the perturbation of the perovskite-type unit in this layered phase, $e_g$ electron density is concentrated randomly in $d_{3x^2-r^2}$ and $d_{3y^2-r^2}$ orbitals [Fig. 13(b)]. At low doping ($0.95 < x < 1.00$), $e_g$ orbitals on $Mn^{3+}$ sites are uncorrelated in $x$ and $y$, giving isolated ferromagnetic units but no net ferromagnetic or Type-C/C* moment. At higher doping ($0.90 < x < 0.95$), $e_g$ orbitals correlate in one dimension to form delocalised bands mediating double exchange along ferromagnetic 'rods' [Fig. 13(c)]. The experimental observation of an orthorhombic structure in this region indicates that these 'rods' lie parallel (along *y*), possibly as a consequence of the constrained 2-D nature of the perovskite layers. Nonetheless, they remain uncorrelated in $x$ and therefore no net ferromagnetic or Type-C/C* reflections are observed. We note in Fig. 3 that as $x$ decreases from 1.00, the saturation moment of the Type-G (or tilted Type-G) magnetic ordering decreases as we approach the Type-C phase boundary from above. In the absence of any additional magnetic reflections over this composition range, this observation suggests that an increasing number of Mn spins are not participating in long-range magnetic ordering. We suggest that this is further evidence in support of the development of uncorrelated Type-C/C* clusters. A similar trend is observed in the Type-C/C* phase field as $x$ approaches the Type-G phase boundary from below. Clearly, focused studies of diffuse scattering, preferably on single crystals, will be required to verify these hypotheses. It should be remembered that the observed tilting of Type-G spins (Table 4) when $x < 1.00$ is an experimental observation rather than a prediction of this qualitative scheme. Nonetheless, the tilt angle $\theta$ does appear to be a useful parameter tracking the development of short-range magnetic order in this mixed-phase regime.

Both Type-C/C* and Type-G magnetic phases were observed in neutron powder diffraction data at $x = 0.90$. As discussed in Section III.C, two alternative models were considered in order to allow the refinement of the magnetic moments due to each phase. In the adopted (magnetic phase segregation) model, as this two-phase regime is approached from the Type-G regime, long-range-ordered Type-C/C* regions grow out of previously non-long-range-ordered regions, rather than out of long-range-ordered Type-G regions; *i.e.*, the total number of spins which are long-range-ordered actually increases. We can now interpret this behaviour in terms of the qualitative scheme for the Type-G regime discussed above (Fig. 13). As $x$ decreases from 1.00, an increasing number of Mn spins do not participate in long-range Type-G magnetic ordering (Fig. 3). For $x < 0.95$, however, these spins appear to be short-range-ordered in the manner of Type-C/C*, as uncorrelated ferromagnetic 'rods' along *b* [Fig. 13(c)]. The magnetic phase segregation model used at $x = 0.90$ now implies that in that two-phase regime, these short-range-ordered 'rods' coalesce into long-range-ordered Type-C/C*, *without affecting the magnetic and crystallographic phase fraction occupied by the long-range-ordered Type-G state*. The net Type-G magnetic moment of the sample is unaffected by this magnetic phase segregation (Figs. 1) and falls on the same $\mu_G$ *vs.* $x$ curve as the rest of the Type-G regime (Fig. 3). That the same behaviour is observed for Type-C/C* moments implies that for $x < 0.90$, short-range-ordered Type-G regions may similarly exist within a Type-C/C* framework. However, since Type-G does not have lower symmetry than Type-C/C* there will be no unambiguous evidence in the lattice parameters. Note that the tilting angle $\theta$ of Type-G spins to the *c* axis, which grows as *x* decreases from 1.00, shrinks again for $x = 0.90$ (Table 4). This, too, is consistent with phase segregation at that composition; the Type-G regions should become more like the ideal Type-G state ($x = 1.00$) again when the Type-C/C* regions coalesce.

# V. CONCLUSIONS

A key motivation for studying mixed-valent manganite perovskites in the R-P series is the potential they offer to investigate relationships among charge, spin and lattice degrees of freedom in a reduced-dimensional setting, and thereby better understand those relationships in 3-D perovskite analogues. The low-temperature magnetic structures of $La_{2-2x}Sr_{1+2x}Mn_2O_7$ in the range $0.50 \le x \le 1.00$ all have direct analogues in mixed-valent manganite perovskites. At high *x* in particular, the Type-C and Type-C/C* regimes occupy approximately the same $Mn^{3+}$-doping (*i.e.* *x*) regions of the phase diagram as they do in $La_{1-x}Ca_xMnO_3$[18, 39] and can be understood in much the same way[38]. These layered phases do indeed yield novel information by virtue of their reduced-dimensional setting, specifically the fact that extended ferromagnetic 'rods' in Type-C/C* are constrained to lie parallel to one another in the *a-b* plane. This demonstrates that the ferromagnetic 'rods' in perovskite Type-C are mediated by double exchange in long-range, pseudo-delocalised, $d_{3z^2-r^2}$ bands. Furthermore, in the layered system it appears that these 'rods' lie parallel even when they are not correlated with one another. They therefore have a structural signature (symmetry lowering) detectable by neutron powder diffraction even in the absence of long-range magnetic ordering. This feature of the layered system provides strong evidence for short-range ordered Type-C units (*i.e.* ferromagnetic 'rods') within a Type-G framework as *x* decreases from 1.00, leading to a phase-segregated (Type-G + Type-C/C*) region before the Type-C/C* regime is entered. Neumeier and Cohn[14] have in fact proposed a very similar phenomenology for the magnetic phase behaviour of $La_{1-x}Ca_xMnO_3$ in the same range of *x*, based on measurements of saturation magnetisation. The findings of the present study provide support for their model by analogy from the 2-D to the 3-D case. Ultimately, all these magnetic ground states follow from the polarisation of $Mn^{3+}$ $e_g$ orbitals which places electron density preferentially along the *y*-axis, initially on isolated sites (0-D) and subsequently in delocalised rods (1-D).

The analogy between $La_{2-2x}Sr_{1+2x}Mn_2O_7$ and $La_{1-x}Ca_xMnO_3$ breaks down at lower values of *x*. In the perovskite case, a long-range CO Type-CE AFI state at $x = 0.50$ persists up to a two-phase region with Type-C[18]. In the present ($n = 2$) case, long-range CO associated with Type-CE (but no actual long-range Type-CE spin ordering) is observed only at $x = 0.50$[5-8, 10], in a narrow temperature range between $T_{CO}$ and $T_N$ (for a Type-A AFI ground state). The present study has found evidence that CO may persist in a thin 'wedge' of *x*-T phase space above $T_N$ for $0.50 \le x \le 0.60$. While many perovskite manganites $Ln_{1-x}Ca_xMnO_3$ exhibit CO when $x = 0.50$[30-36], the relative stability of CO Type-CE AFI, Type-A AFI and FM states around that composition varies significantly.

Given the finely balanced competition among states, and the demonstrated influence of perovskite *A*-site cation radius on the balance between them[22, 23, 30], the only appropriate system with which to compare $La_{2-2x}Sr_{1+2x}Mn_2O_7$ when considering the effects of reduced dimensionality is $La_{1-x}Sr_xMnO_3$[22]. In this system the FM state dominates at low temperature up to $x = 0.54$. For $x \ge 0.54$, the FM state forms below $T_C$ but is replaced below $T_N$ ($< T_C$) by the Type-A AFI state. CO is not observed at all. The key change in a reduced-dimensional setting is that as *x* decreases, Type-A AFI holds off FM as the ground state configuration down to lower values of $x$. This implies that in reduced dimensions, $d_{x^2-y^2}$ orbitals continue to delocalise at lower values of $x$, forming a pseudo-2-D band mediating double-exchange in the *a-b* plane. This band (and hence the Type-A AFI state) is destroyed as $x$ decreases by mounting $Mn^{3+}$ J-T distortions, which lift the degeneracy of the $e_g$ orbitals and stabilise one-dimensional $d_{3z^2-r^2}$ orbitals. [Note that for the perovskite $Nd_{1-x}Sr_xMnO_3$[40] substitution of a smaller *A*-site cation also stabilises the CO Type-CE state by reducing $W$, in agreement with Akimoto *et al.*[23].]

Fig. 10 shows why a higher density of J-T active $Mn^{3+}$ ions (*i.e.* lower $x$) is required for this to occur in the $n = 2$ compared with the perovskite phase. When $x = 1.00$, no $Mn^{3+}$ ions are present and the perovskite phase is cubic, all Mn-O distances being equivalent. In the $n = 2$ phase, this structural 'baseline' at $x = 1.00$ is already distorted by the steric impact of the rock-salt layer on the perovskite layer, elongating the axial bond Mn-O1 compared to the equatorial Mn-O3. (Note that in the layered phase, Mn-O2 is effectively a 'spectator' bond responding to the other Mn-O bond lengths by maintaining an appropriate bond valence sum for the average oxidation state of $Mn$[26].) As discussed in relation to the Type-G and Type-C/C* states, as $x$ decreases, $e_g$ electron density goes into the *a-b* plane and elongates the equatorial relative to the axial bonds. In perovskite, this leads to more distorted $MnO_6$ octahedra as $x$ decreases; in the $n = 2$ phase (Fig. 10) it leads to less distorted $MnO_6$ octahedra as $x$ decreases down to ~ 0.6, at which point they are almost regular. This effectively suppresses the J-T distortion, stabilising the Type-A AFI ground state in the (La,Sr) $n = 2$ manganite to lower $x$ (~ 0.48) than in the perovskite analogue (~ 0.54)[22]. Akimoto *et al* [23] stabilised Type-A down to even lower $x$ (0.40) by substituting Ca for Sr at doping levels above 30 %. This result was similarly interpreted as a consequence of $e_g$ orbital polarisation; calculation of Madelung potentials demonstrated that Ca substitution stabilises $d_{x^2-y^2}$ orbitals relative to $d_{3z^2-r^2}$ orbitals, with a concomitant switch from FM to Type-A AFI states.

Similar effects of perovskite layering can be seen in (Pr,Sr) and (Nd,Sr) manganites. $Pr_{1-x}Sr_xMnO_3$ and $Nd_{1-x}Sr_xMnO_3$ at $x = 0.50$ both have FM states below $T_C$ and AFI states below $T_N < T_C$[41], whereas $Pr_{2-2x}Sr_{1+2x}Mn_2O_7$ and $Nd_{2-2x}Sr_{1+2x}Mn_2O_7$ at $x = 0.50$ both have AFI states below $T_N$ but no FM state[42; 43]. In contrast, it is interesting to note that the spin-glass region found below $T_g$ in the (Gd,Sr), (Dy,Sr), (Ho,Sr) and (Eu,Sr) $x = 0.50$ perovskite manganites are also found in their $n = 2$ analogues at $x = 0.50$[43]; since there is no 2-D delocalised $d_{x^2-y^2}$ band in the spin-glass state, the location of the spin-glass regime with respect to $x$ is independent of $MnO_6$ octahedral distortions and therefore unaffected by reduced dimensionality.

The significance of delocalised $d_{x^2-y^2}$ orbitals to the Type-A AFI state, and the equally important role played by delocalised $d_{3z^2-r^2}$ orbitals in the Type-C/C* state, allows some insight into the presence of a magnetically frustrated regime between the two (Fig. 3). The absence of long-range spin order suggests the absence of such orbital delocalisation, *i.e.* the transition between long-range 1-D delocalised $d_{3z^2-r^2}$ (Type-C/C*) and 2-D delocalised $d_{x^2-y^2}$ (Type-A) bands as $x$ decreases is discontinuous. This differs from the smooth transition from Type-G to Type-A predicted by Maezono and Nagaosa[24], even more than does the presence of a Type-C/C* state. It appears that spin ordering in this system is very closely coupled to $e_g$ orbital polarisation, and hence highly sensitive to distortions of the perovskite unit as discussed above. The same theoretical models, applied to accurate structural data (such as reported in the present study), may well predict a magnetic phase diagram closer to the experimental one described here.

Two distinct effects of reduced dimensionality on the magnetic behaviour of mixed-valent manganite perovskites have been identified in this study. The first of these is that the occupied axial $e_g$ orbitals on $Mn^{3+}$ sites, the delocalisation of which mediates the Type-C/C* AFI state, are constrained to lie in the basal plane of the layered compounds. This allows the transition from ideal Type-G (no $Mn^{3+}$ sites) to Type-C/C* to be more easily observed and understood. It also leads to a magnetically frustrated region between the Type-C/C* and Type-A AFI regimes, presumably due to the incompatibility between 1-D $d_{3y2-r2}$ delocalisation (Type-C/C*) and 2-D $d_{x2-y2}$ delocalisation (Type-A), both in the *a-b* plane. Nonetheless, for $x \geq 0.75$ the relative stability of Type-G and Type-C/C* AFI states is largely determined by the extent of $Mn^{3+}$ doping into the $Mn^{4+}$ lattice, so the magnetic phase diagrams of the 3-D and 2-D phases are very similar in this region. For $x \leq 0.66$, however, reduced dimensionality affects the polarisation of $e_g$ orbitals, and consequently the relative stability of Type-A AFI and FM states, which are mediated by delocalised $d_{x2-y2}$ bands and axial $d_{3z2-r2}$ orbitals respectively. The magnetic phase diagrams of the 2-D and 3-D phases are therefore quite different in this region. Studying layered manganites provides complementary information to the perovskites and can illuminate effects in the latter compounds that are more difficult to observe due to their pseudo-cubic symmetry, thus contributing to our understanding of spin and orbital ordering in mixed valent manganite perovskites.

## AKNOWLEDGEMENTS

Preliminary synchrotron XRD data was collected by D. E. Cox at Brookhaven National Laboratory. This work was supported by the U.S. Department of Energy, Basic Energy Sciences - Materials Sciences, under contract W-31-109-ENG-38.

# APPENDIX

For Type-G, Type-C/C* and Type-G+C/C*, total spin-only moment due to Mn:

$$\mu_{TOT} = \mu_{oG} + \mu_{tG} + \mu_{oC} + \mu_{oD} + \mu_{tD} \tag{1}$$

Subscript notation:    t = tetragonal

o = orthorhombic

G = long-range ordered Type-G

C = long-range ordered Type-C/C*

D = non-long-range ordered

[Note: Long-range ordered Type-C/C* moments can only exist in the orthorhombic phase fraction $\therefore \mu_{tC} = 0$.]

Neutron powder diffraction data includes Type-G and Type-C/C* magnetic reflections (proportional to the squares of the respective magnetic moments), as well as tetragonal and orthorhombic nuclear reflections (note that tetragonal and orthorhombic Type-G magnetic reflections overlap). Rietveld refinement of this data is therefore sensitive to:

observable Type-G moment $\mu_{Gobs} = \mu_{oG} + \mu_{tG}$ (2a)

observable Type-C/C* moment $\mu_{Cobs} = \mu_{oC}$ (2b)

orthorhombic crystallographic phase fraction $F_o$

Type-G and Type-C/C* magnetic moments per Mn site within the crystallographic phase fraction they occupy (note that since Type-G reflections from the orthorhombic and tetragonal crystallographic phase fractions overlap, we must assume $\mu_G$ is the same in both):

$$\mu_G = \mu_{oG}/F_o = \mu_{tG}/(1 - F_o) \tag{3a}$$
$$\mu_{C/C*} = \mu_{oC}/F_o \tag{3b}$$

## A. Type-G ($0.90 < x \leq 1.00$)

No Type-C/C* reflections $\therefore \mu_{oC} = 0$

$\therefore$ from (1)        $\mu_{TOT} = \mu_{oG} + \mu_{tG} + \mu_{oD} + \mu_{tD}$

$\therefore$ from (2a)        $\mu_{Gobs} = \mu_{oG} + \mu_{tG} = \mu_{TOT} - \mu_{oD} - \mu_{tD}$ (4)

$\therefore$ from (3a)        $= \mu_G.F_o + \mu_G.(1 - F_o)$

$\mu_G = \mu_{Gobs}$ (5)

## B. Type-C/C* ($0.75 \leq x < 0.90$)

No Type-G reflections $\therefore \mu_{oG} = \mu_{tG} = 0$

$\therefore$ from (1)        $\mu_{TOT} = \mu_{oC} + \mu_{oD} + \mu_{tD}$

$\therefore$ from (2b)        $\mu_{Cobs} = \mu_{oC} = \mu_{TOT} - \mu_{oD} - \mu_{tD}$ (6)

$\therefore$ from (3b)        $= \mu_{C/C*}.F_o$

$\mu_{C/C*} = \mu_{Cobs}/F_o$ (7)

## C. Type-G + Type-C/C* (*x* = 0.90)

<u>Chemical Phase Segregation Model</u>: Long-range ordered Type-C/C* moments and long-range ordered Type-G moments exist in *different* orthorhombic crystallites within the polycrystalline sample, which we denote by the superscripts C and G respectively on the orthorhombic terms, *i.e.* $\mu_{oG} = {}^{G}\mu_{oG}$, $\mu_{oC} = {}^{C}\mu_{oC}$, $\mu_{oD} = {}^{G}\mu_{oD} + {}^{C}\mu_{oD}$ and $F_{o} = {}^{G}F_{o} + {}^{C}F_{o}$.

$$\therefore \text{ from (1)} \qquad \mu_{TOT} = {}^{G}\mu_{oG} + \mu_{tG} + {}^{C}\mu_{oC} + {}^{G}\mu_{oD} + {}^{C}\mu_{oD} + \mu_{tD}$$

$$\therefore \text{ from (2a,b)} \qquad \mu_{Gobs} + \mu_{Cobs} = {}^{G}\mu_{oG} + \mu_{tG} + {}^{C}\mu_{oC} = \mu_{TOT} - \mu_{oD} - \mu_{tD} \qquad (8)$$

$$\therefore \text{ from (3a,b)} \qquad = \mu_{G}.{}^{G}F_{o} + \mu_{G}.(1 - {}^{G}F_{o} - {}^{C}F_{o}) + \mu_{C/C*}.{}^{C}F_{o}$$

$$\mu_{G}.(1 - {}^{C}F_{o}) + \mu_{C/C*}.{}^{C}F_{o} = \mu_{Gobs} + \mu_{Cobs} \qquad (9)$$

Eqn. 9 cannot be solved by refinement without fixing an additional variable since we cannot determine ${}^{C}F_{o}$, *e.g.*

$$\text{fix} \qquad \mu = \mu_{G} = \mu_{C/C*}$$

$$\therefore \text{ from (9)} \qquad \mu = \mu_{Gobs} + \mu_{Cobs}$$

(10)

<u>Magnetic Phase Segregation Model</u>: Long-range ordered Type-C/C* moments and long-range ordered Type-G moments exist in *the same* orthorhombic crystallites within the polycrystalline sample.

$$\therefore \text{ from (2a)} \qquad \mu_{Gobs} = \mu_{oG} + \mu_{tG} = \mu_{TOT} - \mu_{oC} - \mu_{oD} - \mu_{tD} \qquad (11a)$$

$$\therefore \text{ from (3b)} \qquad = \mu_{G}.F_{o} + \mu_{G}.(1 - F_{o})$$

$$\mu_{G} = \mu_{Gobs} \qquad (12a)$$

$$\& \text{ from (2b)} \qquad \mu_{Cobs} = \mu_{oC} = \mu_{TOT} - \mu_{oG} - \mu_{tG} - \mu_{tD} - \mu_{oD} \qquad (11b)$$

$$\therefore \text{ from (3a)} \qquad = \mu_{C/C*}.F_{o}$$

$$\mu_{C/C*} = \mu_{Cobs}/F_{o} \qquad (12b)$$

TABLE 1. FWHM of synchrotron XRD reflections at 300 K. $\lambda$ = 0.688 Å, step size 0.005 ˚$2\theta$.

| $x$ | FWHM (0, 0, 10) (˚$2\theta$) | FWHM (2 0 0) (˚$2\theta$) |
|------|------|------|
| 1.00 | 0.015(5) | 0.020(5) |
| 0.95 | 0.025(5) | 0.030(5) |
| 0.90 | 0.025(5) | 0.050(5) |
| 0.80 | 0.025(5) | 0.050(5) |
| 0.75 | 0.025(5) | 0.045(5) |
| 0.55 | 0.030(5) | 0.025(5) |
| 0.50 | 0.025(5) | 0.025(5) |

TABLE 2. Data from final Rietveld-refinements of La$_{2-2x}$Sr$_{1+2x}$Mn$_2$O$_7$ at various values of $x$ using GPPD ($x$ = 0.58) and SEPD data.

|  | $x$ = 0.58 (20 K) | $x$ = 0.80 (300 K) | $x$ = 0.80 (10 K) | $x$ = 0.92 (20 K) | $x$ = 0.98 (20 K) |
|---|---|---|---|---|---|
| Symmetry | *I*4/*mmm* | *I*4/*mmm* | *Immm* | *Immm* | *I*4/*mmm* |
| $a$ (Å) | 3.853224(6) | 3.83470(7) | 3.78894(5) | 3.797285(9) | 3.800450(13) |
| $b$(Å) | - | - | 3.86404(5) | 3.802282(9) | - |
| $c$(Å) | 19.8686(9) | 20.0037(4) | 19.9629(2) | 20.0010(13) | 20.1006(19) |
| Mn $z$ | 0.0971(2) | 0.09666(17) | 0.09715(11) | 0.0972(3) | 0.0974(5) |
| LaSr2 $z$ | 0.31797(12) | 0.31674(9) | 0.31681(6) | 0.3163(2) | 0.3161(3) |
| O2 $z$ | 0.19361(16) | 0.19276(12) | 0.19340(8) | 0.1924(3) | 0.1925(4) |
| O3 $z$ | 0.09498(12) | 0.09515(9) | 0.09502(13) | 0.09534(13)[†] | 0.09581(19) |
| O4 $z$ | - | - | 0.09551(13) | 0.09534(13)[†] | - |
| Mn-O1 (Å) | 1.930(5) | 1.934(3) | 1.939(2) | 1.944(7) | 1.957(10) |
| Mn-O2 (Å) | 1.917(6) | 1.922(4) | 1.922(3) | 1.905(9) | 1.913(13) |
| Mn-O3 (Å) | 1.92708(13) | 1.91758(7) | 1.93249(8) | 1.90150(15) | 1.90049(19) |
| Mn-O4 (Å) | - | - | 1.89475(6) | 1.89900(15) | - |
| $R_p$ | 0.0483 | 0.0569 | 0.0387 | 0.0551 | 0.0762 |
| $wR_p$ | 0.0661 | 0.0903 | 0.0549 | 0.0766 | 0.1212 |
| $R(F^2)$ | 0.0495 | 0.0778 | 0.0485 | 0.0738 | 0.0710 |
| $\chi^2$ | 2.612 | 1.471 | 5.040 | 2.117 | 1.104 |
| Spin State | A AFI | - | C/C* AFI | G AFI | G AFI |
| $\mu$ ($\mu_B$/Mn) | 2.32(5) | - | 1.85(4) | 1.67(4) | 2.40(4) |

Atomic positions: in *I*4/*mmm* (# 139) Mn in 4*e*, LaSr1 in 2*b*, LaSr2 in 4*e*, O1 in 2*a*, O2 in 4*e*, O3 in 8*g*; in *Immm* (# 71) Mn in 4*i*, LaSr1 in 2*c*, LaSr2 in 4*i*, O1 in 2*a*, O2 in 4*i*, O3 in 4*j*, O4 in 4*j*.

[†],Constrained to be equal.

TABLE 3. Evolution of nuclear and magnetic phase fractions (PF) and magnetic moments across the Type-C/C* AFI regime of $La_{2-2x}Sr_{1+2x}Mn_2O_7$ at 20 K (Rietveld refined SEPD data).

| $x$ | Nuclear PF $Immm/(Immm+I4/mmm)$ | Magnetic PF C*/(C+C*) | $\mu_{C/C*}$ ($\mu_B$/Mn) |
|---|---|---|---|
| 0.75 | 0[†] | 1[†] | 1.24(8) |
| 0.78 | 0.612(2) | 1[†] | 2.04(8) |
| 0.80 | 0.931(16) | 1[†] | 1.85(4) |
| 0.82 | 0.937(16) | 0.775(16) | 1.78(4) |
| 0.84 | 0.982(16) | 0.807(14) | 1.78(2) |
| 0.86 | 0.983(16) | 0.69(10) | 1.28(12) |
| 0.88 | 0.953(18) | 0.81(11) | 1.28(8) |
| 0.90 | 0.955(4) | 1[†] | 0.94(5)[‡] |

[†] Fixed

[‡] Using a magnetic phase segregation model (see Section III.C)

TABLE 4. Evolution of spin magnitudes and orientations ($\theta$ = angle to $z$) across the Type-G AFI regime at 20 K (Rietveld refined SEPD data). See Section III.D for discussion of azimuthal angles.

| $x$ | $\theta$ (°) | $\mu$ ($\mu_B$/Mn) |
|---|---|---|
| 0.90 | 42(12) | 1.22(7)[†] |
| 0.92 | 68(10) | 1.67(3) |
| 0.94 | 68(5) | 1.98(3) |
| 0.96 | 60(2) | 2.21(2) |
| 0.98 | 24(5) | 2.40(4) |
| 1.00 | 8(8) | 2.39(3) |

[†] Using a magnetic phase segregation model (see Section III.C)

FIG. 1. Magnetic and crystallographic phase diagram of La$_{2-2x}$Sr$_{1+2x}$Mn$_2$O$_7$. Solid data points represent magnetic transitions determined from neutron powder diffraction data (antiferromagnetic Néel temperature T$_N$, ferromagnetic Curie temperature T$_C$, charge-ordering temperature T$_{CO}$ and orbital ordering temperature T$_O$). Open data points represent crystallographic transitions. Lines are guides to the eye.

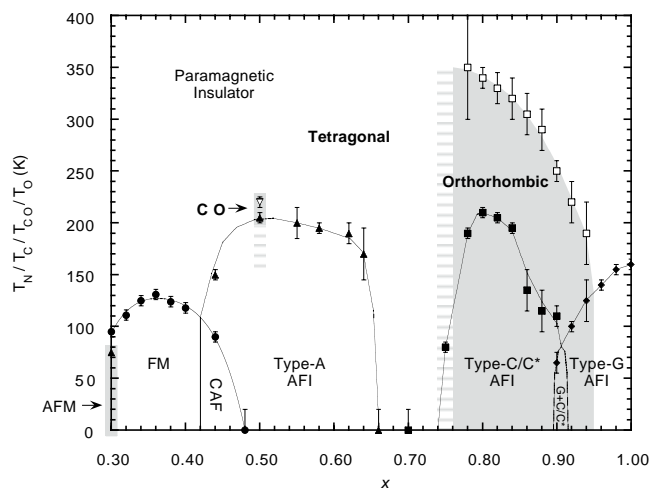

FIG. 2. Schematic illustrations of magnetic structures of La$_{2-2x}$Sr$_{1+2x}$Mn$_2$O$_7$, arrows representing spin orientations on Mn sites: (a) AFM, (b) FM at $x = 0.32$, (c) FM at $x = 0.40$, (d) Type-A AFI, (e) Type-C AFI, (f) Type-C* AFI, (g) Type-G AFI at $x = 0.92$ and (h) Type-G AFI at $x = 1.00$.

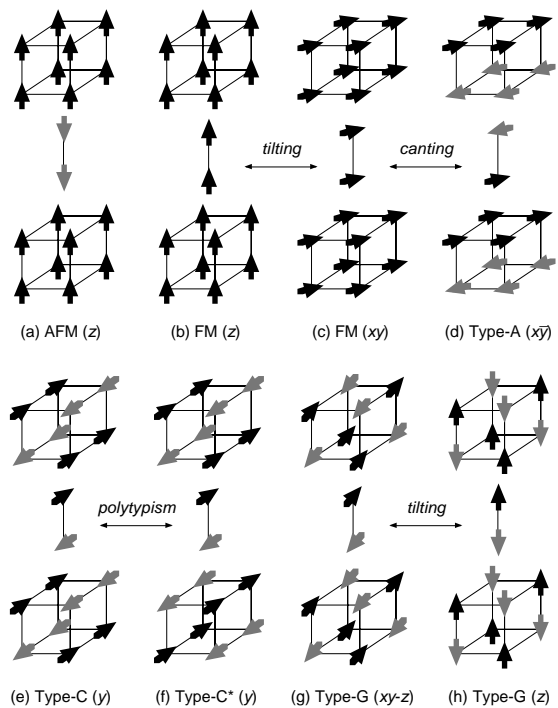

FIG. 3. Rietveld-refined magnetic moments at 20 K (Type-A and Type-G) or 10 K (Type-C/C*) for the various states of overdoped La$_{2-2x}$Sr$_{1+2x}$Mn$_2$O$_7$, compared to the total expected spin-only moment. At $x = 0.90$ two different models have been used to explain the observation of both Type-G and Type-C/C* moments; the open marker represents chemical phase segregation model and the grey markers represents the magnetic phase segregation model (see Section III.C). The orbitals (or delocalised bands) occupied by Mn$^{3+}$ $e_g$ electrons are indicated. Lines are guides to the eye.

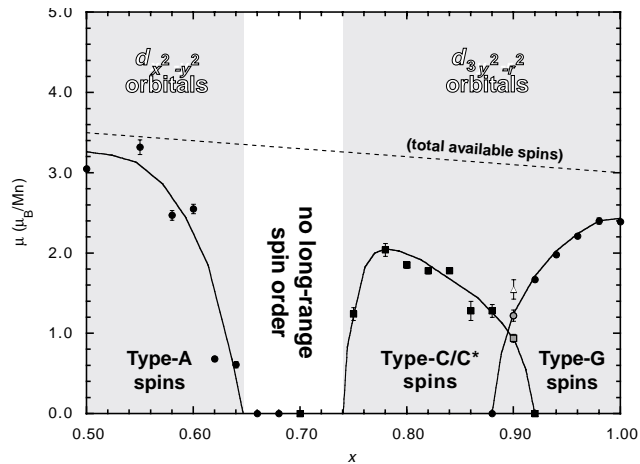

FIG. 4. Observed (.), calculated and difference (below) plots of Rietveld-refined GPPD neutron powder diffraction data (90 ° detector bank) for La$_{2-2x}$Sr$_{1+2x}$Mn$_2$O$_7$, $x = 0.58$ at 20 K. The bottom row of reflection markers refers to the $I4/mmm$ nuclear structure and the top row to the Type-A AFI magnetic structure. Some prominent Type-A magnetic peaks are indexed, a perovskite (*) and an $n = 1$ R-P (†) impurity peak are marked.

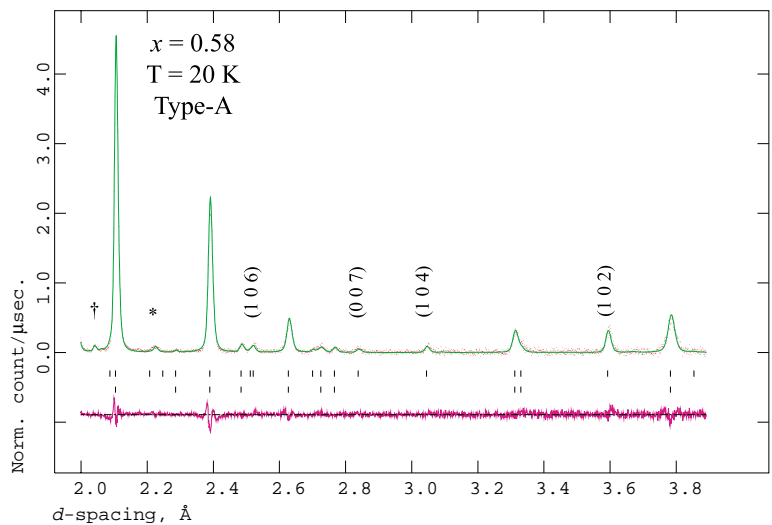

FIG. 5. Observed (.), calculated and difference (below) plots of Rietveld-refined SEPD neutron powder diffraction data (60 ˚ detector bank) for La$_{2-2x}$Sr$_{1+2x}$Mn$_2$O$_7$, $x = 0.84$ at 10 K. The bottom row of reflection markers refers to the *I*4/*mmm* nuclear structure, the middle row to the *Immm* nuclear structure and the top row to the Type-C/C* AFI magnetic structures. Some prominent Type-C ($h + 1/2$, $k$, $l$) and Type-C* ($h + 1/2$, $k$, $l + 1/2$) magnetic peaks are indexed. Inset shows the splitting of the (2 0 0) reflection in the middle of the tetragonal-orthorhombic transition upon cooling.

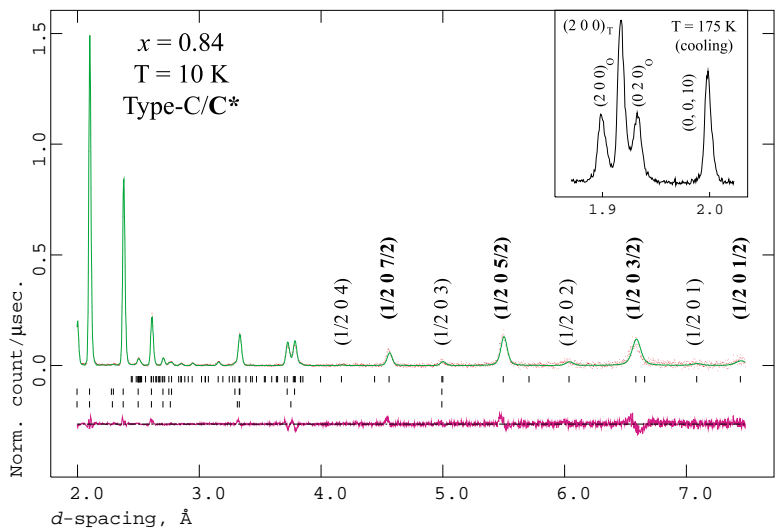

FIG. 6. (a) *I4/mmm* nuclear phase fraction *vs.* T for La$_{2-2x}$Sr$_{1+2x}$Mn$_2$O$_7$, $x = 0.80$ on cooling (empty circles) and warming (filled circles), and Type-C* AFI magnetic moment (assuming Type-C* exists in both *Immm* and *I4/mmm* nuclear phases) *vs.* T on cooling (empty squares) and warming (filled squares). (b) Type-C* AFI magnetic moment (assuming Type-C* exists in only the *Immm* nuclear phase) *vs.* T on cooling (empty squares) and warming (filled squares). Lines are guides to the eye (from Rietveld refinement of SEPD neutron powder diffraction data).

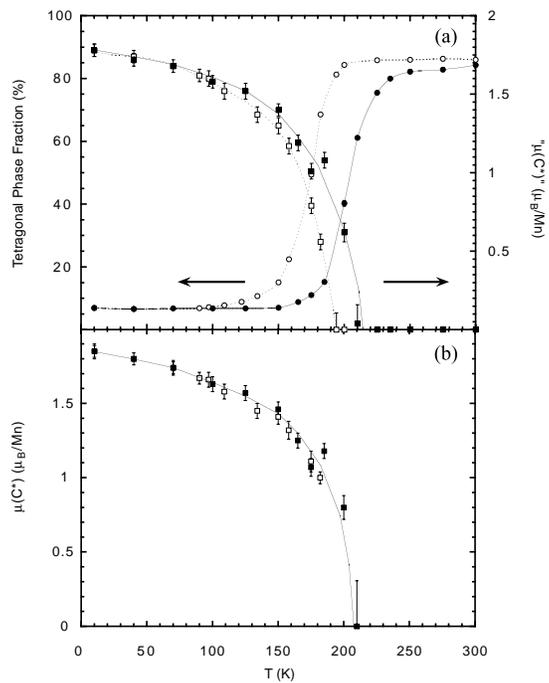

FIG. 7. Unit cell dimensions *vs. x* at room temperature (filled circles) and low temperature ($5 \leq T \leq 20$ K) (empty circles) (from Rietveld refinement of GPPD, SEPD and HIPD neutron powder diffraction data). Error bars are smaller than symbols. Lines are guides to the eye.

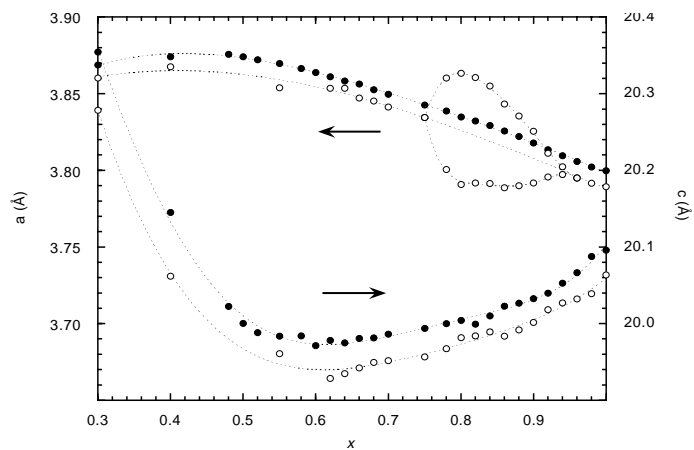

FIG. 8. Magnetic moments due to the Type-C* and Type-G states *vs.* T at *x* = 0.90, normalised in each case to the moments at 10 K. Lines are guides to the eye.

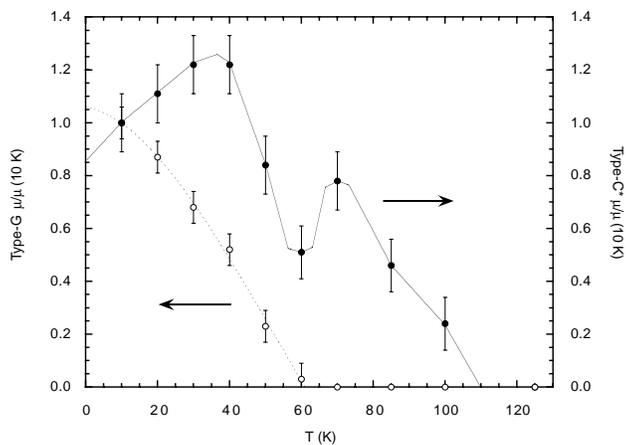

FIG. 9. Observed (.), calculated and difference (below) plots of Rietveld-refined SEPD neutron powder diffraction data (60 ° detector bank) for La$_{2-2x}$Sr$_{1+2x}$Mn$_2$O$_7$, (a) $x = 0.94$ and (b) $x = 0.98$ at 20 K. The bottom rows of reflection markers refers to the *I4/mmm* nuclear structures and the top rows to the Type-G AFI magnetic structures. $\theta$ is the angle to $z$ by which the Type-G spins are tilted. Some prominent magnetic peaks are indexed and a perovskite impurity peak is marked (*).

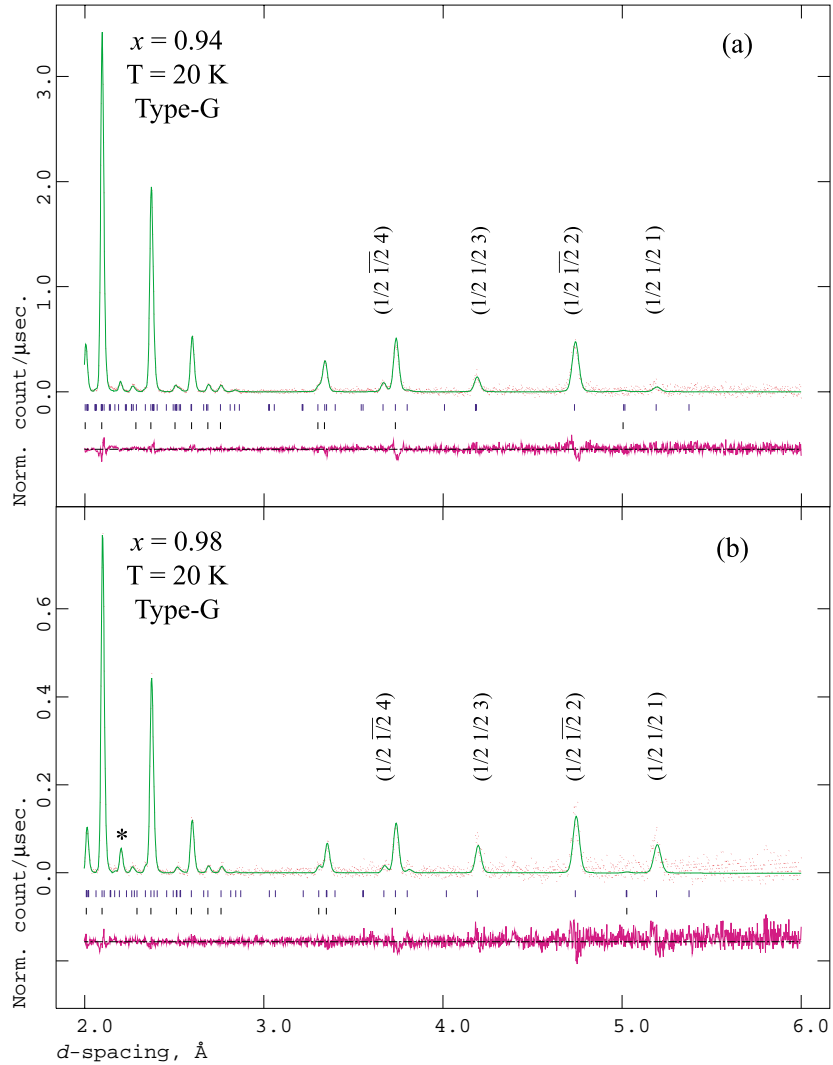

FIG. 10. MnO$_6$ octahedral bond lengths *vs. x* at room temperature (from Rietveld refinement of GPPD, SEPD and HIPD neutron powder diffraction data). Lines are guides to the eye.

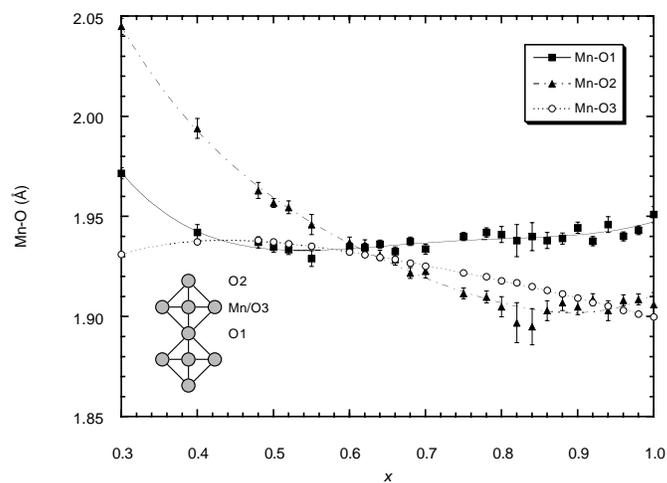

FIG. 11. Lattice parameters *vs.* T for (a) $x = 0.55$, (b) $x = 0.58$, (c) $x = 0.80$ and (d) $x = 0.96$ (from Rietveld refinement of HIPD (a), GPPD (b) and SEPD (c, d) neutron powder diffraction data). Lines are guides to the eye.

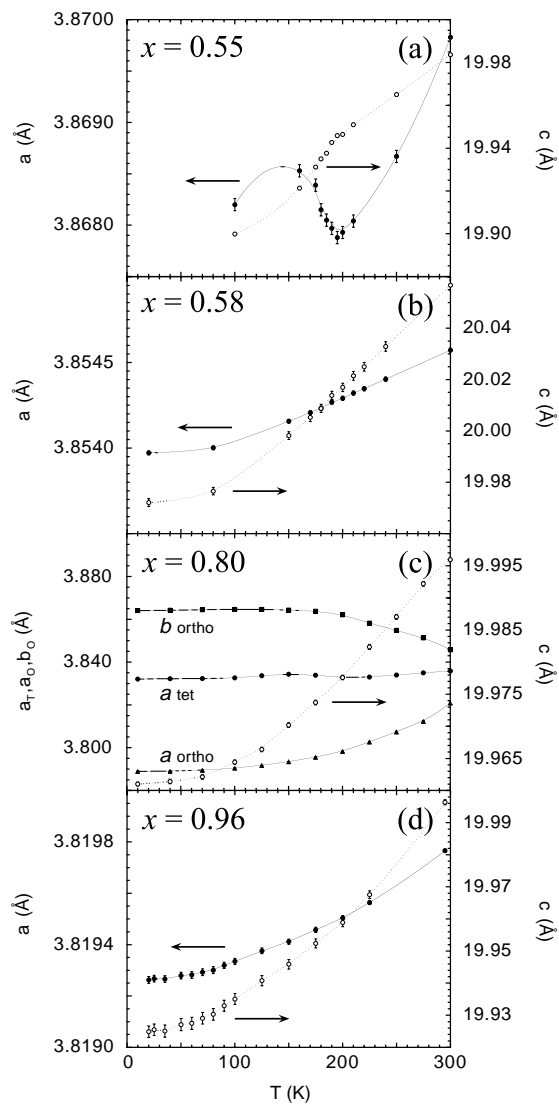

FIG. 12. Normalised resistances of pellets *vs.* T for various values of *x* within the Type-A AFI regime.

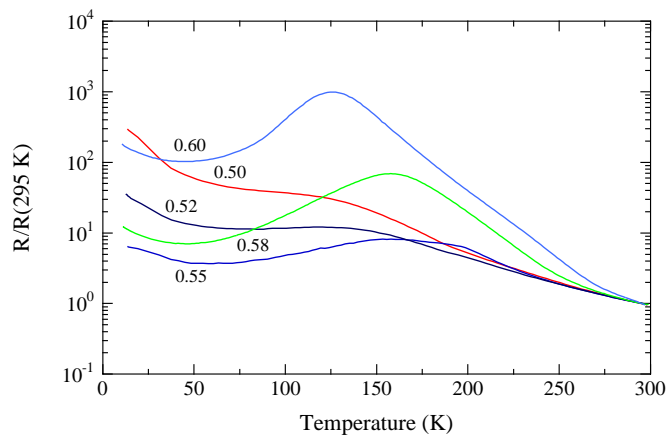

FIG. 13. Schematic illustration of spins on Mn sites in the *a-b* plane evolving from the Type-G to the Type-C/C* AFI states. Dark grey $d_{3y2-r2}$ orbitals are intended to indicate the density of $Mn^{3+}$ sites rather than exact ordering patterns; pseudo-delocalised $d_{3y2-r2}$ bands are light grey. (a) No $Mn^{3+}$. (b) $Mn^{3+}$ uncorrelated along *a* and *b*. (c) $Mn^{3+}$ uncorrelated along *a* but correlated along *b*. (d) $Mn^{3+}$ correlated along *a* and *b*.

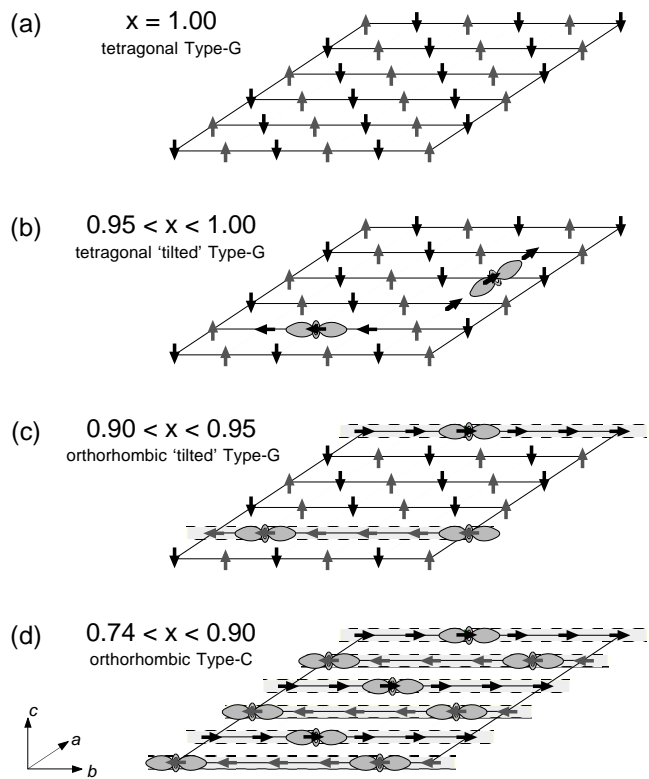